\documentclass[showpacs,amssymb,floatfix,pre,aps,notitlepage,twocolumn]{revtex4-1}

\usepackage{amssymb,amsfonts,amsmath,bm}
\usepackage{graphics,graphicx}
\usepackage{color}

\newcommand{\sinc}{\hbox{sinc}}

\newcommand{\bk}{{\bm k}}
\newcommand{\beq}{\begin{equation}}
\newcommand{\eeq}{\end{equation}}
\newcommand{\emma}[1]{#1}

\begin{document}



\title{When random walkers help solving intriguing integrals}

\author{Satya N. Majumdar and Emmanuel Trizac} 
\affiliation{
LPTMS, CNRS, Univ. Paris-Sud, Universit\'e Paris-Saclay, 91405 Orsay,
France }
\date{\today} 

\begin{abstract}
We revisit a family of integrals that delude intuition, and that recently
appeared in mathematical
literature in connection with computer algebra package verification.
We show that the remarkable properties displayed by these integrals become transparent 
when formulated in the language of random walks. In turn, the random walk view naturally 
leads to a plethora of nontrivial generalizations, that are worked out. Related 
\emma{complex} identities 
are also derived, without the need of explicit calculation.
\emma{The crux of our treatment lies in a causality argument where a message that travels at finite speed 
signals the existence of a boundary.}
\end{abstract}

\maketitle

{\em Introduction}. While intuitions and experimentations are both crucial in 
mathematical works, 
inductive thinking may be spectacularly misguided in some cases. 
A celebrated illustration of the dangers of pattern extrapolation is
provided by the question of circle division by chords \cite{Pizza}: Consider $n$
points on the circumference of a circle and join every pair of points by a chord
such that at any point inside the circle at most two chords can intersect.
How many regions $S_n$ gets the circle divided into? By simple drawing, one sees
that $S_1=1$ (by convention), $S_2=2$, $S_3=4$, $S_4=8$, $S_5=16$. At
this point one may naively guess that for general $n$, $S_n=2^{n-1}$. Wrong !
It turns out that $S_6=31$. Indeed, the correct answer is $S_n= {n\choose 4}+ 
{n\choose 2}+1$, which happens to coincide with the sequence $2^{n-1}$ up to $n=5$,
but starts differing from it for $n=6$ onwards ! 

Our interest goes here to a lesser known such problem, and the surprising behavior of
integrals of the type
\begin{eqnarray}
I_N &=& \int_{-\infty}^{\infty} \prod_{n=1}^N \sinc\left(\frac{k}{2n-1}\right) \, dk  
\label{eq:typea}\\
J_N &=& \int_{-\infty}^{\infty} \cos(k) \, \prod_{n=1}^N \sinc\left(\frac{k}{2n-1}\right) \, dk
\label{eq:typeb}
\end{eqnarray}
where $\sinc(x)=\sin(x)/x$ denotes the cardinal sine function \cite{rque1,rque2}. 
\emma{We do not dwell on the prevalence
of $\sinc$ function in mathematics (geometry, spectral analysis\ldots) 
and physics (signal processing, optics\ldots),
see e.g. \cite{GS}.}
It was shown that $I_1=I_2=I_3=I_4=I_5=I_6 = I_7= \pi$, 
whereas $I_N < \pi $, for all $N\geq 8$ \cite{BoBo01}. In the latter situation, the difference 
$\pi-I_N$ is minute, less than $10^{-10}$ for $N=8$, which was first realized 
numerically,
and attributed to a bug in the software \cite{BoBo01}. A related phenomenon was observed for
the $J$-family: $J_N = \pi/2$ for $N\leq 56$, but $J_N < \pi/2$ for all $N\geq 57$
\cite{Schm14}. A theorem shown in \cite{BoBo01} rationalizes this matter of fact:
it states that provided $\sum_{n=2}^N |a_n| \, < \, |a_1|$,
\beq
\frac{1}{2\pi}\, \int_{-\infty}^\infty \prod_{n=1}^N \sinc(a_n\, k) \, dk \, = 
\, \frac{1}{2|a_1|} .
\label{eq:statement}
\eeq
Without loss of generality, one can \emma{choose} the coefficients $a_n$ to be positive real quantities,
and $a_1$ can then be \emma{taken} as the largest of them. Given that $\sum_{n=2}^7 1/(2n-1)<1$
while $\sum_{n=2}^8 1/(2n-1)>1$, this explains the behavior of the $I$-family,
for which $a_1=1$. 
The $J$-family  falls under the same argument \cite{rque10}. 
When the equality in \eqref{eq:statement} breaks, the explicit integrals could be computed.
The corresponding values, related to the volume of hypercubes, cut by parallel hyperplanes,
is immaterial for our purposes \cite{BoBo01}. Our goal is rather to provide a transparent
understanding of the statement \eqref{eq:statement}. To this end, we will show that 
the language of random walks, \emma{and physical intuition}
not only provide a natural framework to understand this change of behavior, but also 
leads to relevant and interesting generalizations, thereby
offering an explicit and effort-free calculation of complex multidimensional 
integrals \cite{comment52}. \emma{At the heart of our approach lies a {\em causality} argument,
formulated in terms of a message that signals the existence of a boundary.}

{\em Random walkers in a finite or infinite ``world''}.
We start by considering a random walk making $N$ steps on a line,  
$x_{N} = \sum_{n=1}^N \eta_n$ starting at $x_0=0$, where $\eta_n$ is uniformly 
distributed in 
$[-a_n,a_n]$ and the $\eta_n$'s are independent random variables. 
The probability density function (pdf) of each $\eta_n$ is thus a rectangle function,
with simple characteristic function (Fourier-Transform)
$\langle e^{i k \eta_n} \rangle \,=\, \sinc(a_n\, k)$.
The characteristic function of $x_N$, sum of independent increments, thus reads
\beq
\Big\langle e^{i k x_N} \Big\rangle  \, =\, \prod_{n=1}^N \Big\langle e^{i k \eta_n} 
\Big\rangle \,\, =\,\, \prod_{n=1}^N \sinc(a_n\, k)
\eeq
which allows us to write its pdf as the inverse Fourier transform
\begin{eqnarray}
&&p_N(x_N) = \int_{-\infty}^\infty \, \frac{dk}{2\pi} \, \prod_{n=1}^N \sinc(a_n\, k) \, e^{-i k x_N} \label{eq:FTpos}\\
\Longrightarrow ~~ &&\frac{1}{2\pi}\, \int_{-\infty}^\infty \prod_{n=1}^N \sinc(a_n\, k) \, dk \, = p_N(0).
\label{eq:FT}
\end{eqnarray}
The $I_N$ integral under scrutiny is thus isomorphic to $p_N(0)$, i.e., the 
probability 
density of the random walk to be back at the origin, 
while starting from the origin ($x_0=0$).

\begin{figure}[t!]
    \begin{center}
        \includegraphics[width=2.8cm]{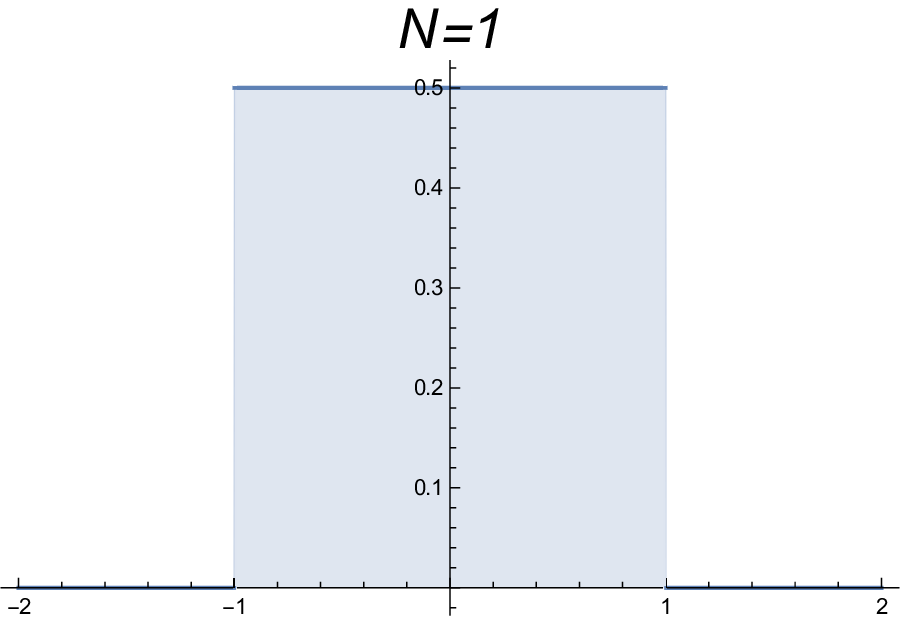}
        \includegraphics[width=2.8cm]{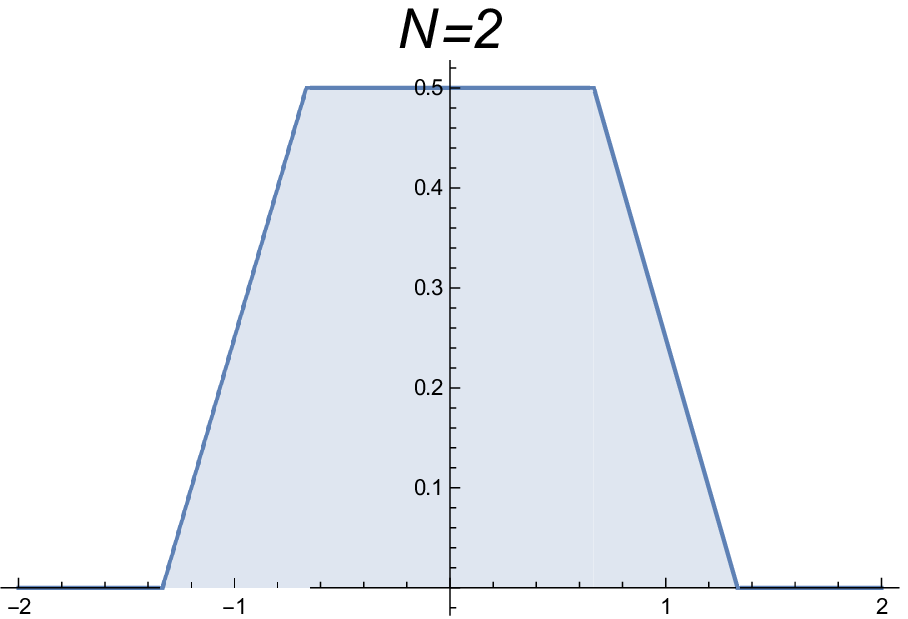}
        \includegraphics[width=2.8cm]{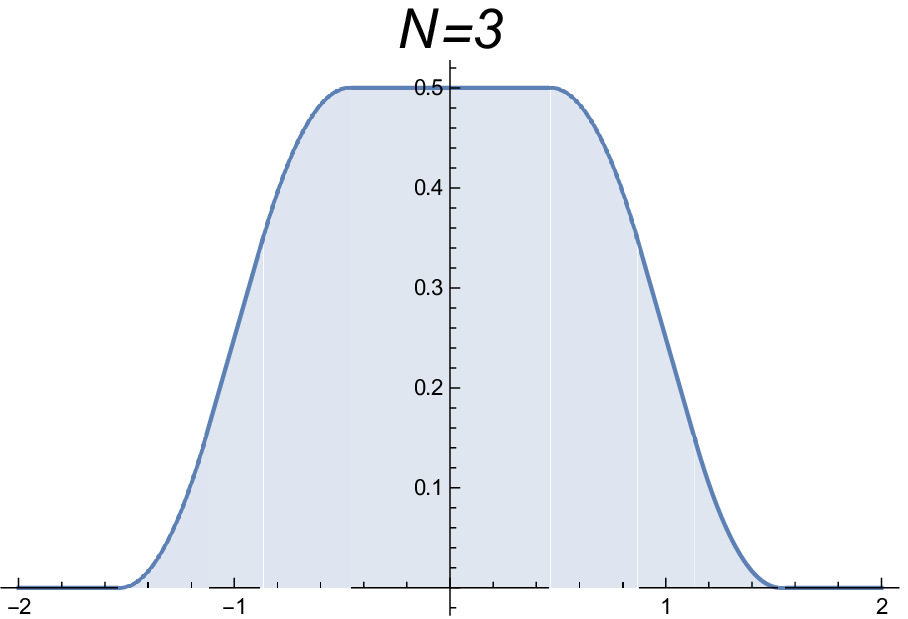}
	\caption{Probability density function of the random walk $x_N$ after 
	$N=1$, $N=2$ and $N=3$ steps, as indicated. Here, the amplitudes of the steps are 
	$a_1=1$, $a_2=1/3$, $a_3=1/5$, in line with the definition of $I$ integrals in Eq. 
	\eqref{eq:typea}. The density at the origin is invariant
	(equal to 1/2), which means that $I_1=I_2=I_3=\pi$. For respectively 
	$N=1,2$ and 3, the extension of the flat 
	region near the origin is $2a_1$, $2(a_1-a_2)$, $2(a_1-a_2-a_3)$, due to the progression 
	of the walkers arriving from the boundaries (so-called ``messengers'' in the main text).
	Note that the density at $x=1$ is preserved as well, and half that at the origin (leaving aside the 
	$N=1$ case). }
	\label{fig:pdf123}
    \end{center} 
\end{figure}

To proceed further, it is useful to reinterpret $p_N(0)$ as follows.
Consider a large (infinite actually) number of independent random walkers, all
starting at $x_0=0$. Then $p_N(0)$ is just the fraction of walkers at $x=0$ after
$N$ steps, i.e., the density of this gas of independent particles at the origin
after $N$ steps. 
After step 1, their density is uniform in $[-a_1,a_1]$ so that $p_1(0)=1/(2 a_1)$ 
(incidentally meaning that
$I_1=\pi$).  A second step is then made, with amplitude $a_2<a_1$. 
Because the jump $a_2$ is finite and $a_2<a_1$, it is clear that all the walkers
that were, after step 1, in the range $[-(a_1-a_2), a_1-a_2]$ will not leave
their step-1 domain $[-a_1,a_1]$ following the second step. Only walkers near
the two edges, e.g., those in the range $[a_1-a_2,a_1]$ or $[-a_1, -a_1+a_2]$ may leave the 
step-1 domain after the second jump. Hence, the walkers in $[-(a_1-a_2), a_1-a_2]$
do not `see' the edges of the step-1 domain--for them, it is as if the system was infinite
with uniform density $1/(2a_1)$.    
In such an ``infinite world'',
the gain and loss contribution balancing those walkers leaving the origin and 
those reaching it after the second step do cancel: $p_1(0)=p_2(0)$. This is 
illustrated 
in Fig. \ref{fig:pdf123} where one can appreciate that the flatness of the density
near the origin is preserved, although in a range that diminishes with 
the number $N$ of steps
performed. The argument does not depend on the left/right symmetry of the random steps 
\cite{rque65}. \emma{In other words, we invoke causality and the boundedness of the steps to
state that 
the only possibility for $p_N(0)$}
to be affected by a new step is when walkers having started from the edges
at $\pm a_1$ do reach the origin. Those `messengers' carry the information 
that 
the ``world'' is not infinite, which in turn impinges on $p_N(0)$. 
If $\sum_{n=2}^N a_n < a_1$, the distance traveled by the
messengers is not sufficient to reach the origin, and $p_N(0)=1/{2a_1}$ is 
$N$-independent.
We therefore recover statement \eqref{eq:statement} \cite{rque66}.
\emma{Besides, while our random walk argument directly applies to $I_N$ integrals, it also is relevant for 
the $J_N$-type, as explained in the supplemental material \cite{suppl}. It is nevertheless necessary here
to supplement the analysis with a new property, the left-right symmetry of the random steps.
The key feature becomes the preservation of the edge density $p_N(a_1)$ under performing random steps, while 
it pertained to the preservation of $p_N(0)$ when treating $I_N$ integrals \cite{suppl}.}

The random walk reformulation provides us with an immediate
generalization. Consider, for instance, the case where the first step
is of Pearson's
type \cite{Pearson} (i.e. of a fixed amplitude $a_1$, ending at $\pm a_1$), while the 
subsequent steps are again uniform as before with $\eta_n\in [-a_n, a_n]$ for $n\ge 
2$,
then $p_N(0)=0$ for $\sum_{n=2}^N a_n< a_1$.
In this case, the characteristic function after $N$ steps is given by:
$\langle e^{ik\, x_n}\rangle= \cos(k\, a_1)\, \prod_{n=2}^N \sinc(a_n k)$.
\emma{Then our causality argument tells us that}
$p_N(0)=0$ for $\sum_{n=2}^N a_n< a_1$.
Evidently, the origin 
remains void of walkers, until the messengers arrive at $x=0$.  Provided
that $\sum_{n=2}^N a_n \, < \, a_1$,
this implies 
\cite{rque50}:  
\beq
\int_{-\infty}^\infty \cos(a_1 k)\prod_{n=2}^N \sinc(a_n\, k) \, dk \, = 
\,0 .
\label{eq:cos}
\eeq
\emma{This identity can be recovered by invoking a different random walk sharing
with the previous one steps $2,3\ldots n$, but not the first step \cite{suppl}.}
    
{\em One dimensional generalizations}.
A natural extension of the above results consists in considering that all steps 
except the first are arbitrary, but of finite range. The corresponding pdfs are therefore
of finite support. The argument now involves the associated characteristic functions, 
that we denote $\widehat{\cal F}_n(k)$. These $\widehat{\cal F}_n(k)$ are defined as the Fourier
transforms of pdfs ${\cal F}_n(x)$ that have a finite support, taken for convenience
to be unity.
Provided $\sum_{n=2}^N a_n \, < \, a_1$, one can write:
\beq
\frac{1}{2\pi}\, \int_{-\infty}^\infty \sinc(a_1 k) 
\prod_{n=2}^N \widehat{\cal F}_n(a_n\, k) \, dk \, = 
\, \frac{1}{2 a_1} .
\label{eq:statement_general}
\eeq
\emma{Under the same condition and 
taking once more advantage of causality,  we obtain \cite{rque50}}
\beq
\int_{-\infty}^\infty \cos(a_1 k)  \prod_{n=2}^N \widehat{\cal F}_n(a_n\, k) \, dk \, = 
\, 0
\label{eq:statement_general_cos}
\eeq
where again all amplitudes $a_n$ are considered positive. Some 
particular cases have been addressed in earlier studies \cite{BoBo01,AlGu14},
but we stress that many more are subsumed under Eqs. \eqref{eq:statement_general}
and \eqref{eq:statement_general_cos}. The task  amounts to establishing a catalogue 
of eligible $\widehat{\cal F}_n(k)$. It is not our purpose here, and we simply
mention some emblematic such functions: the Bessel functions $J_0(k)$, $J_1(k)/k$ and more
generally $J_\nu(k)/k^\nu$ for $\nu>-1/2$, $(1-\cos k)/k^2$, $(k-\sin k)/k^3$
and a number of hyper-geometric functions. To generate candidates, advantage can be taken 
from the study of hyper-uniform systems, that feature potentials of bounded Fourier Transform,
see e.g. \cite{Torquato}.

{\em Beyond dimension one}. 
A second natural extention of previous considerations 
consist in considering $d$ dimensional random walks, with
$d>1$. 
A straightforward calculation shows that the counterparts of the one 
dimensional 
$\sinc(ka)$ and $\cos(ka)$ functions are given as follows. For a
one-step walk with jump $\bm \eta$, chosen respectively (i) uniformly {\em within} a 
$d$-dimensional sphere 
of radius $a$ and (ii) uniformly {\em on} the surface of the same sphere (Pearson's 
type jump), the
associated characteristic functions of the jumps are
\begin{eqnarray}
\langle e^{i\, {\bm k} \cdot {\bm \eta} }\rangle= 
\begin{cases}
& \frac{(2\pi)^{d/2}}{V_d}\, 
\frac{J_{d/2}(ka)}{(ka)^{d/2}}\quad\quad\,\, {\rm (i)} \\
&\\ 
& 
\frac{(2\pi)^{d/2}}{S_d}\,
\frac{J_{d/2-1}(ka)}{(ka)^{d/2-1}} \quad\quad {\rm (ii)}
\end{cases}
\label{jump.d}
\end{eqnarray}
where $V_d= \pi^{d/2}/{\Gamma(d/2+1)}$ and $S_d=d\, V_d$ are respectively
the volume and surface of a $d$-dimensional unit sphere. For $d=1$,
using $J_{1/2}(z)= \sqrt{2/{\pi\, z}}\, \sin(z)$ and
$J_{-1/2}(z)= \sqrt{2/{\pi z}}\, \cos(z)$, one recovers
respectively $\sinc(ka)$ and $\cos(ka)$.
A new catalogue of functions $\widehat{\cal F}_n^{(d)}(\bk)$ can then 
be established, such that their $d$-dimensional 
Fourier Transform is of bounded support (several interesting candidates can also
be found in  \cite{Torquato}; a rather generic one being the hyper-geometric function 
$_1F_2\left( \frac{d+m}{2};\frac{d}{2};1+\frac{m+d}{2},-k^2  \right)$, where $m$ is some arbitrary parameter). 
Knowing the eligible building blocks $\widehat{\cal F}_n^{(d)}(\bk)$, one can write
upon setting $k=|\bk|$ 
\beq
 \int_{\mathbb{R}^d} \frac{J_{d/2}(a_1 k)}{k^{d/2}} \,
\prod_{n=2}^N \widehat{\cal F}_n^{(d)}(a_n\, \bk) \, d^d \bk \, = 
\left(\frac{2\pi}{a_1}\right)^{d/2},
\label{eq:statement_general_ddim}
\eeq
provided $\sum_{n=2}^N a_n \, < \, a_1$.

A nontrivial identity
follows from Eq. 
(\ref{eq:statement_general_ddim})
by considering a special case. Choose the $n$-th step uniformly from
a $d_n$-dimensional sphere of radius $b_n$ (for $n\ge 1$).
Consequently, using $\widehat{\cal F}_n^{(d_n)}(a_n\, \bk)\propto 
\emma{J_{d_n/2}(b_n k)/k^{d_n/2}}$ in Eq. (\ref{eq:statement_general_ddim}), we get upon setting 
$d_n=2 \mu_n$ the following identity 
\begin{eqnarray}
\int_0^\infty &&  dk \, k^{\mu_1-1} \, J_{\mu_1}(b_1\,k) \,\prod_{j=2}^N 
\frac{J_{\mu_j}(b_j k)}{k^{\mu_j}} \nonumber\\
&&=  \frac{2^{\mu_1-1-\sum_{j=2}^N \mu_j}}{{b^{\mu_1}_1}} \,
\frac{\Gamma(\mu_1)}{\prod_{j=2}^N \Gamma(1+\mu_j)}  \,  
\prod_{j=2}^N b_j^{\mu_j} 
\end{eqnarray}
provided $\sum_{j=2}^N b_j \,<\, b_1$ \cite{rque101}. 
Not surprisingly, with a Pearson first step that depopulates the origin
and for  $\sum_{n=2}^N a_n \, < \, a_1$:
\beq
 \int_{\mathbb{R}^d} \frac{J_{d/2-1}(a_1 k)}{k^{d/2-1}} \,
\prod_{n=2}^N \widehat{\cal F}_n^{(d)}(a_n\, \bk) \, d^d \bk \, = \, 0 .
\label{eq:statement_general_ddim_pearson}
\eeq
Mixing dimensions, we note that the steps $n\geq 2$ can be of any type provided the
associated Fourier Transform is bounded: 
in \eqref{eq:statement_general_ddim} and \eqref{eq:statement_general_ddim_pearson}, the building 
blocks  $\widehat{\cal F}_n^{(d)}(\bk)$ can be some $\widehat{\cal F}_n^{(d')}(\bk)$, borrowed 
from a lower dimensional catalogue with $d'<d$.
In doing so, we generate a wealth of complex integrals. Some of the simplest 
are known \cite{Grads,Wats}, for instance
\beq
\int_0^\infty k^{\nu-\mu+1}\, J_{\nu} (a_1 k) \, J_\mu{(a_2 k)} \, \cos (a_3 k) 
\,\sinc(a_4 k) \, = 0  \, 
\eeq
for $a_2+a_3+a_4<a_1$, which follows from \eqref{eq:statement_general_ddim_pearson} with $\nu=d/2-1$, $\widehat{\cal F}_{n=2}^{(d)}(\bk)=J_\mu(k)/k^\mu$
and that appears under section 6.711.2 in \cite {Grads}, when $a_4=0$ \cite{rque20}.
Yet, infinitely many other identities that are subsumed 
in \eqref{eq:statement_general_ddim} or \eqref{eq:statement_general_ddim_pearson} 
are complex and seemingly unknown.

\begin{figure}[t!]
    \begin{center}
     \includegraphics[width=4cm]{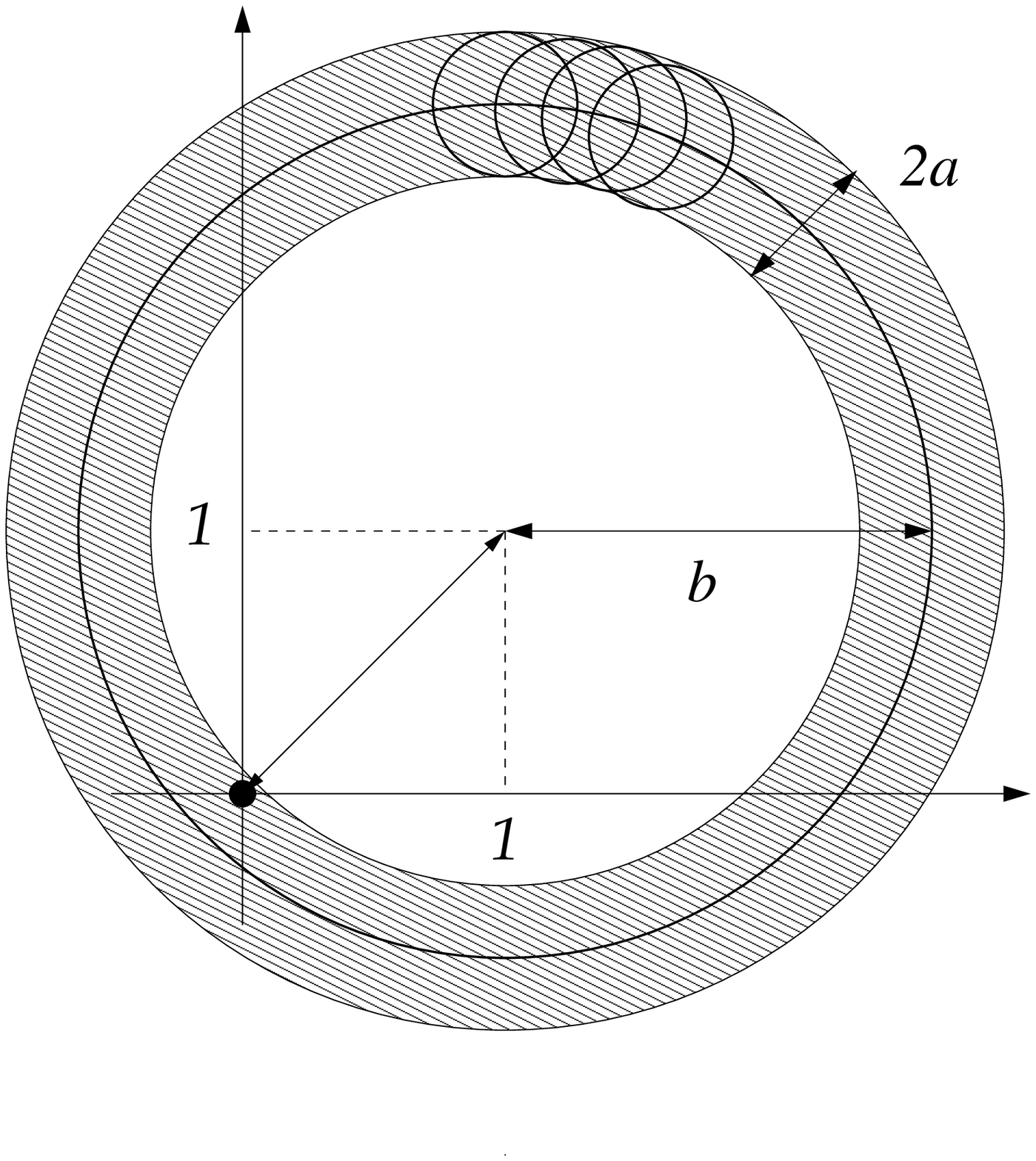}
	\caption{Sketch of the random walk geometry considered. After step 2, the walker is on one of four corners
	of a square, chosen to be $(1,1)$ on the graph. The third step is uniform on 
the large circle with radius $b$,
	while the fourth is uniform on a smaller circle with radius a. The walker can then end up, non uniformly,
	at any point within the shaded annulus of inner radius $b-a$ and outer radius $b+a$. 
	Since the origin lies in that region, the integral considered in non-vanishing.}
	\label{fig:sketch}
    \end{center} 
\end{figure}

It is worth stressing that in some cases the random walk reformulation may not 
immediately lead to an explicit result, however it may nevertheless offer a 
direct means of calculation. As an example, we find the following nontrivial
identity
\begin{eqnarray}
\int_{\mathbb{R}^2} && d k_x \, dk_y\, \cos(k_x) \cos(k_y)  \, J_0(b k)\, J_0(a k) \nonumber\\ 
&& = \frac{4}{\sqrt{\left(\left(a+b\right)^2-2\right)\,\left(2-\left(a-b\right)^2\right)}}
\label{fourstep.1}
\end{eqnarray}
with 
$k=\sqrt{k_x^2+k_y^2}$. The above result holds for $\sqrt{2} \in 
[|b-a|,b+a]$ (and say $a>0$, $b>0$); otherwise, the 
integral vanishes. 
\emma{The proof is provided in the supplemental material \cite{suppl}. We outline here the main steps.} Consider
a $2$-d random walk, starting at the origin ${\bm 0}$ and making $4$ successive steps: 
a first jump $\pm 1$ along $x$-direction, a second jump
$\pm 1$ along $y$-direction, then a Pearson jump on the circle of radius $b$ with a
final fourth step
distributed as the third, but with a radius $a$ and say $a<b$, see Fig. \ref{fig:sketch}.
Then the lhs of Eq. (\ref{fourstep.1}),
using the results from line (ii) of Eq. (\ref{jump.d}) ($d=1$ for the first two steps
and $d=2$ for the last two steps),
is precisely $4\pi^2\, p_4(\bm 0)$ 
where $p_4(\bm 0)$ denotes the density at the origin ${\bm 0}$ after $4$ steps.
This density can, in turn be computed by elementary means, see the Suppl. Mat. \cite{suppl},
leading to the rhs of Eq. (\ref{fourstep.1}). 
If $\sqrt{2} \notin [b-a,b+a]$, the random walk which is at a distance $\sqrt{2}$ from the
origin after step 2 cannot be back to be origin after step four, exploring an annulus with inner radius $b-a$ and outer radius $b+a$
(see Fig. \ref{fig:sketch}).
Equally tractable is the case where the fourth step is not 
Pearson but uniform within the disc of radius $a$ (for the 4-th step we
use (i) of Eq. (\ref{jump.d}) with $d=2$), leading to
\begin{eqnarray}
\int_{\mathbb{R}^2} && d k_x \, dk_y\, \cos(k_x) \cos(k_y) \, 
J_0(b k) \, \frac{J_1(a k)}{k} \nonumber\\ 
&& = 
\frac{2}{a}\, \arccos\left(\frac{2+b^2-a^2}{2\sqrt{2}\,b}\right)\, . 
\label{fourstep.2}
\end{eqnarray}
The lhs can be made more complex without sacrificing the 
possibility of an explicit calculation of $p_4(\bm 0)$.

{\em When sums and integrals coincide}.
We now turn to a distinct problem, that bears a similarity with the previous ones, after 
suitable reformulation. There was some interest recently in identities of the form \cite{BaBB08,DoGL12}
\beq
\int_{-\infty}^\infty \prod_{n=1}^N \sinc(a_n k) \, dk = \sum_{k=-\infty}^\infty \prod_{n=1}^N \sinc(a_n k) 
\label{eq:sum=integral}
\eeq
which hold provided $\sum_{n=1}^N a_n < 2\pi$ \cite{rque30}. The latter condition can be compared to that applying to \eqref{eq:statement},
$\sum_{n=2}^N a_n <  a_1$,
that can be rewritten as $\sum_{n=1}^N a_n < 2 a_1$. The difference between the two criteria, where the same 
quantity if bounded either by $2\pi$ or by $2 a_1$,
indicates that the identity \eqref{eq:sum=integral} cannot be reduced to any of the previous arguments. Yet, the random walk reformulation 
also is insightful to show, and understand, relation \eqref{eq:sum=integral}. The idea is to compare two population of random walkers, one on the infinite line (case F,
for ``flat''),
and the other on the unit circle (case C). Both population, starting from the origin, undergo the same random jumps.
Provided that the front-runners (the random walkers having travelled the \emma{greater} distance from the origin)
did not travel round the circle in case C, moving on a flat line or on a finite circle is immaterial.
The corresponding condition reads $\sum_{n=1}^N a_n < 2\pi$. When this inequality is fulfilled,
the \emma{probability density} of the walkers at the origin is thus the same in cases C and F.
Expressing the pdf as either a Fourier series for case C or a Fourier transform for case F, we then get
\beq
\int_{-\infty}^\infty \prod_{n=1}^N \widehat{\cal F}_n(a_n k) \, dk = \sum_{k=-\infty}^\infty \prod_{n=1}^N \widehat{\cal F}_n(a_n k) ,
\label{eq:flatearth}
\eeq
which includes \eqref{eq:sum=integral} and many other cognate relations \cite{Stormer}.
As above, the $\widehat{\cal F}_n$ refer to arbitrary functions, the Fourier transform of which are bounded with unit support.
Loosely speaking, Eq. \eqref{eq:flatearth} can thus be viewed as the ``flat world equation''.

\vskip 0.2cm

{\em Conclusion and discussion}. We have proposed a random walk interpretation of a curious phenomenon,
exhibited by integrals of type \eqref{eq:typea}-\eqref{eq:typeb}. The underlying physical image is that 
of an ensemble of random walkers starting from the origin, and 
performing a first step so as to populate uniformly the interval $[-a_1,a_1]$.
The walkers then undergo
a series of $N$ smaller steps with respective amplitudes $a_n$. If the maximal span of these steps 
cannot lead walkers from the edge (i.e. at $x=\pm a_1$) back to the origin $x=0$, 
then the walkers near $x=0$ have a fixed density (given by the lhs in Eq. 
\eqref{eq:statement}), specified by the first step and thus equal to $1/(2 a_1)$.
In pictorial terms, the walkers near the origin cannot know they live in a finite world,
unless the messengers starting from the confines at $\pm a_1$ reach them.
This may never happen if $\sum_{n=2}^\infty a_n \, < \, a_1$, in which case 
an equality like \eqref{eq:statement} will hold at all orders $N$ \cite{rque60}.
\emma{It is interesting here to note that the model of random walks with shrinking steps directly applies
to physico-chemical problems such as line broadening for single molecule spectroscopy in disordered media
\cite{Barkai,PaulK}}.
The random walk reformulation naturally leads to non-trivial extensions, 
since it is irrelevant that the
steps $n=2,3,$ etc. be uniformly distributed, provided the first one (labeled $n=1$) has the desired
property (uniform to lead to \eqref{eq:statement} or Pearson to lead to \eqref{eq:cos})
and that the subsequent steps ($n=1, 2$ \ldots) are bounded. 
Generalizations in higher dimensions appear of particular interest, and 
provide calculation-free results that would otherwise require considerable effort and ingenuity.



\emma{While the mathematical problem at stake deceives intuition, we have shown that physical arguments may take over.
Physics' insight takes the form of a causality rule, where a message travels from a boundary. 
Applied to random walkers undergoing jumps of bounded amplitude, this yields a clear account for the change of behavior
of a class of multidimensional integrals (of the types \eqref{eq:statement_general_ddim} and 
 \eqref{eq:statement_general_ddim_pearson}).  
 The random walk picture also allows for simple calculation of complex multidimensional integrals (such as in \eqref{fourstep.1}) and 
\eqref{fourstep.2}). Besides, related probabilistic ideas can be generalized to compute other classes of
integrals that are otherwise hard to obtain. Assume for instance that a 1d walker starts
with a Pearson jump of amplitude $\pm a_1$, so that $\langle |x_1| \rangle =a_1$. 
The boundedness of subsequent steps and the causality rule mean that $\langle |x_N| \rangle$ remain at  $a_1$,
as long as $\sum_{n=2}^N a_n \leq a_1$, i.e. as long as the message starting from $x=a_1$ 
does not hit $x=0$. This implies 
\begin{equation}
\int_0^\infty \frac{1}{k^2} \, \left(1-\cos{(k a_1)}  \prod_{n=2}^N \widehat{\cal F}_n(a_n k) \right) \, dk \,=\, 
\frac{\pi}{2} \, a_1
\end{equation}
after a straightforward re-expression of $\langle |x_N| \rangle$ presented in \cite{suppl}, section III.
It thus appears that the curious phenomenon at work for Borwein integrals is much more general
and applies to a much broader class of complex integrals and discrete sums.
}

%



\begin{thebibliography}{99}
\bibitem{rque10}
Noticing that $\cos(k) \,\sinc(k) = \sinc(2k)$, one should here
consider $a_1=2$ instead of $a_1=1$. Then, $N=56$ is the threshold such that 
$\sum_{n=2}^{56} 1/(2n-1)<2$
while $\sum_{n=2}^{57} 1/(2n-1)>2$.
\bibitem{comment52}
\emma{Our approach shares with previous ones \cite{BoBo01,Schm14,AlGu14} the explicit usage 
of Fourier transforms. Yet, it differs in spirit in its probabilistic 
perspective.}
\bibitem{rque65}
\emma{A Fokker-Planck equation establishes this fact, but it
proves relevant for future extensions to remain at a more intuitive level of understanding.}
\bibitem{rque66}
\emma{We reiterate here that the walk is commutative, as much as the products appearing 
in \eqref{eq:typea}; we single out the first step ($n=1$) as being that with the largest
amplitude among the $a_n$'s.}
\bibitem{rque50}
Instead of a cosine, we could use any function having a Fourier-Transform void of k component
for $-1<k<1$. such as $(\cos(k)-k\sin(k))/(1-k^2)$.
\bibitem{rque101}
Note that for $d=1$,
using $J_{1/2}(z)=\sqrt{2/{\pi z}}\, \sin(z)$, we recover the 
result in Eq. (\ref{eq:statement}).
\bibitem{rque30}
Taking for the $a_n$ the successive ordered prime numbers $a_1=2$, $a_2=3$ etc.,
one reaches the spectacular conclusion that the equality in Eq. \eqref{eq:sum=integral}
holds for $N<10^{176}$, and breaks down for still larger $N$ \cite{BaBB08}.
\bibitem{rque60}
\emma{With the same reasoning, if step $n=1$ is of Pearson type (leading to $\pm a_1$), then the density 
of walkers at the origin vanishes, as long as the cumulated walk cannot lead from $\pm a_1$
back to $x=0$. This is the very meaning of Eq. \eqref{eq:cos} and its associated condition.}
\end{thebibliography}
\end{document}